\documentclass[a4paper]{jpconf}
\usepackage{graphicx}

\begin{document}
\title{Quantum features of natural cellular automata}

\author{Hans-Thomas Elze}

\address{Dipartimento di Fisica ``Enrico Fermi'', Universit\`a di Pisa,  
Largo Pontecorvo 3, I-56127 Pisa, Italia}

\ead{elze@df.unipi.it} 

\begin{abstract}
Cellular automata can show well known features of quantum mechanics, such as a 
linear rule according to which they evolve and which resembles a discretized version of 
the Schr\"odinger equation. This includes corresponding conservation laws. 
The class of ``natural'' Hamiltonian cellular automata is  
based exclusively on integer-valued variables and couplings and their dynamics derives from 
an Action Principle. They can be mapped reversibly to continuum models by applying Sampling Theory. Thus, ``deformed'' quantum mechanical models 
with a finite discreteness scale $l$ are obtained, which for $l\rightarrow 0$ 
reproduce familiar continuum results. We have recently demonstrated that such 
automata can form ``multipartite'' systems consistently with the tensor 
product structures of nonrelativistic many-body quantum mechanics, while interacting 
and maintaining the linear evolution.  
Consequently, the Superposition Principle fully applies  
for such primitive discrete deterministic automata and their composites and   
can produce the essential quantum effects of interference and entanglement. 
\end{abstract}

\section{Why cellular automata?}
The  {\it Cellular Automaton Interpretation} of Quantum Mechanics (QM) has recently 
been outlined by G.~'t\,Hooft \cite{tHooft2014}.  
This presents an attempt to redesign the foundations of quantum theory in accordance 
with essentially classical concepts, such as determinism and existence of ontological states 
of reality. Which entails a surprising and intuitive explanation of the Born rule and  
the apparent collapse of quantum mechanical states  in measurement 
processes.   

Such an approach may be founded on the observation that quantum 
mechanical features arise in a large variety of deterministic and, loosely speaking, 
``classical'' models.  -- So far, however,  
practically all of these models have been exceptional in that they cannot easily be 
generalized to cover real phenomena, incorporating interactions and relativity. 
Yet Cellular Automata (CA) may provide the necessary versatility, as we shall presently continue to discuss \cite{PRA2014,EmQM13}. For a large variety of earlier 
attempts in this field, see, for example,  
Refs.\,\cite{H1,H2,H3,Kleinert,Elze,Groessing,Khrennikov,Margolus,Jizba,Mairi,Isidro,DAriano} 
and  further references there. 

To begin with, let us recall the linearity of quantum mechanics (QM) as a crucial aspect 
of the unitary dynamics 
embodied in the Schr\"odinger equation. This linearity does not depend on the 
particular object  under study, provided it is sufficiently isolated from 
anything else. It is reflected in the Superposition Principle and implies the possibilities    
of interference effects and of non-factorizable states 
of composite objects, {\it i.e.} entanglement in multipartite systems.    

The linearity of QM has been questioned and nonlinear modifications 
have been proposed before -- notwithstanding various `no-go' arguments -- especially in 
order to test experimentally the robustness of QM against such nonlinear deformations.   
Until now no deviations from the predictions of QM have been observed, in particular no 
indications of nonlinearity. 

We have studied a seemingly unrelated {\it discrete} 
dynamical theory, {\it i.e.}, which appears to deviate substantially from quantum theory. Nevertheless, it has been shown with the help of Sampling Theory 
that the deterministic mechanics of the class of discrete {\it Hamiltonian CA} can be 
mapped one-to-one to continuum models pertaining to nonrelativistic QM, which are  
modified by the presence of a fundamental time scale \cite{PRA2014,EmQM13}.  

Perhaps surprisingly, this construction of a linear relation between CA 
and QM with a nonzero discreteness scale, compatible with 
the consistency of the Action Principle underlying the discrete dynamics on one side 
and the locality of the continuum description on the other, requires that both theories 
are linear \cite{Discrete14}.  
  
This result has led us to consider composite objects formed 
from CA subsystems \cite{Elze16}.  -- Clearly, QM is special in that it is characterized not 
only by interference 
effects, like any classical wave theory would be, but also by the tensor product structures of states and observables applying for composite systems, which allow for entanglement. -- 
It is not obvious that CA can constitute composites which conform with QM in this aspect, 
in the limit of negligible discreteness scale. 
A main obstruction could have been the fact that the state space of Hamiltonian CA is not 
a complex projective space, since the norm of the analogue of state vectors is not 
conserved by the dynamics; 
instead there is a conserved two-time correlation function, as we shall recall, which 
becomes the familiar norm only in the continuum limit.   

In Section\,2., we will review the earlier results concerning individual CA 
and briefly indicate  the construction of multipartite CA --  
in such a way that the Superposition Principle is respected and the composition 
rules of CA are compatible with those of QM. This `outlandish' perspective based on CA 
may lead to additional insight in regard to interference and entanglement and eventually 
to an understanding of nonrelativistic QM in accordance with Ref.\,\cite{tHooft2014}.  
Concluding remarks are presented in Section\,3.  

\section{Natural Hamiltonian CA -- action, evolution, conservation laws and composites}
We describe classical Hamiltonian CA with countably many degrees 
of freedom presently in terms of {\it complex integer-valued} state variables 
$\psi_n^\alpha$ (also known as {\it Gaussian integers}), where 
$\alpha\in {\mathbf N_0}$ denote different degrees of freedom and $n\in {\mathbf Z}$  
different states labelled by this discrete {\it clock variable}.  
Various equivalent forms of the action for such CA exist, as indicated  
earlier \cite{PRA2014}. We will employ a particularly compact form 
here, which is useful for the following construction of composite CA 
in analogy with multipartite QM systems.   
   
Let $\hat H:=\{  H^{\alpha\beta}\}$ denote a self-adjoint matrix of Gaussian integers 
that will play the role of the Hamilton operator. Furthermore, we introduce 
the suggestive notation $\dot O_n:=O_{n+1}-O_{n-1}$, for any quantity $O_n$ depending 
on the clock variable $n$. The  summation convention for Greek indices, 
$r^\alpha s^\alpha\equiv\sum_\alpha r^\alpha s^\alpha$, will often allow us to 
simplify  notation further by suppressing them altogether, for example, writing 
$\psi_n^{*\alpha}H^{\alpha\beta}\psi_n^\beta\equiv\psi_n^*\hat H\psi_n$. 

Then, with $\psi_n^\alpha$ and $\psi_n^{*\alpha}$ as independent variables, 
the CA action ${\cal S}$ is defined by:  
\begin{equation}\label{action} 
{\cal S}[\psi ,\psi^*]\; :=\;
\sum_n\big [\frac{1}{2i}(\psi_n^*\dot\psi_n-\dot\psi_n^*\psi_n)+\psi_n^*\hat H\psi_n\big ] 
\;\equiv\;\psi^*\hat{\cal S}\psi 
\;\;, \end{equation} 
where the operator $\hat{\cal S}$ is a useful abbreviation, {\it cf.} below. 
In order to set up the variational principle, we introduce    
{\it integer-valued} variations $\delta f$ to be applied to a polynomial $g$ 
as follows: 
\begin{equation}\label{variation} 
\delta_{f}g(f):=[g(f+\delta f)-g(f-\delta f)]/2\delta f 
\;\; , \end{equation} 
and $\delta_fg\equiv 0$, if $\delta f=0$. -- 
Note that variations  of terms that are 
{\it constant, linear, or quadratic} in integer-valued variables  yield analogous results as  
standard infinitesimal variations of corresponding expressions in the continuum.  --  
Making use of these ingredients, we postulate the variational principle:      
\vskip 0.15cm \noindent 
[{\it CA Action Principle}] \hskip 0.15cm   
The discrete evolution of a CA is determined by stationarity of the  
action under arbitrary integer-valued variations of all 
dynamical variables, $\delta {\cal S}=0$.\hfill $\bullet$ \vskip 0.15cm  

Let us point out several characteristics of this {\it CA Action Principle}:  
\\ \noindent  
{\bf i)} While infinitesimal variations do not conform with integer valuedness, 
there is {\it a priori} no restriction of integer variations. Hence {\it arbitrary}  
integer-valued variations must be admitted. \\ \noindent 
{\bf ii)} One could imagine contributions to the action (\ref{action})  which are of 
higher than second order in $\psi_n$ or $\psi_n^*$. 
However, in view of arbitrary  variations 
$\delta \psi_n^\alpha$ and $\delta \psi_n^{*\alpha}$, 
such additional contributions to the action 
must be absent for consistency.  
Otherwise the number of equations of motion 
generated by variation of the action, according to Eq.\,(\ref{variation}), 
would exceed the number of variables. (A   
limited number of such remainder terms, which are nonzero only for some fixed values 
of $n$, could serve to encode  
the {\it initial conditions} for the evolution.)    

These features of the {\it CA Action Principle} 
are essential in constructing a map between Hamiltonian CA and equivalent quantum mechanical continuum models \cite{PRA2014}. -- For curiosity, generalizations of the 
variations defined in Eq.\,(\ref{variation}) have also been considered, which allow  
higher than second order polynomial terms in the action. While leading to consistent 
discrete equations of motion, however, these nonlinear equations generally are beset 
with undesirable nonlocal features in the corresponding continuum description 
\cite{Discrete14}.    

\subsection{Equations of motion} 
The equations of motion are obtained by applying the 
 {\it CA Action Principle} to the action ${\cal S}$ of Eq.\,(\ref{action}) with 
the variations as defined in Eq.\,(\ref{variation}). Thus, variations  
$\delta\psi_n^*$ and $\delta\psi_n$, respectively, yield 
discrete analogues of the Schr\"odinger equation and its adjoint: 
\begin{eqnarray}\label{delpsistar} 
\dot\psi_n&=&\frac{1}{i}\hat H\psi_n 
\;\;, \\ [1ex] \label{delpsi} 
\dot\psi_n^*&=&-\frac{1}{i}(\hat H\psi_n)^* 
\;\;, \end{eqnarray} 
recalling that $\hat H=\hat H^\dagger$ and $\dot\psi_n =\psi_{n+1}-\psi_{n-1}$, {\it etc.} 
We remark that the action ${\cal S}$ vanishes when 
evaluated for solutions of these finite difference equations. 

Note that by setting $\psi_n^\alpha =:x_n^\alpha +ip_n^\alpha$, with real 
integer-valued variables $x_n^\alpha$ and $p_n^\alpha$, and suitably separating real and imaginary parts of Eqs.\,(\ref{delpsistar})--(\ref{delpsi}), the discrete equations 
assume a form that resembles Hamilton's equations for a network of coupled 
classical oscillators \cite{Heslot85,Skinner13}:   
\begin{equation}\label{xdotCA} 
\dot x_n^\alpha\;=\;h_S^{\alpha\beta}p_n^\beta +h_A^{\alpha\beta}x_n^\beta  
\;\;,\;  
\;\;\dot p_n^\alpha \;=\;-h_S^{\alpha\beta}x_n^\beta +h_A^{\alpha\beta}p_n^\beta  
\;\;, \end{equation}
where we also split the self-adjoint matrix $\hat H$ 
into real integer-valued symmetric and antisymmetric parts, respectively, 
$H^{\alpha\beta}=:h_S^{\alpha\beta}+ih_A^{\alpha\beta}$. These equations 
led us to the name {\it Hamiltonian CA}; this is furthermore justified by the fact that  analogues of Poisson brackets and classical like observables can be introduced here 
\cite{WignerSymp13}. 

\subsection{Conservation laws} 
The time-reversal invariant equations of motion of Section\,2.1.  
give rise to conservation laws which are in {\it one-to-one correspondence} with those of 
the related Schr\"odinger equation in the continuum. It is straightforward to demonstrate 
the following theorem. 

\vskip 0.15cm \noindent 
[{\it Theorem\,A}] \hskip 0.15cm For any matrix $\hat G$ that commutes with 
$\hat H$, $[\hat G,\hat H]=0$, there 
is a {\it discrete conservation law}: 
\begin{equation}\label{Gconserv} 
 \psi_n^{\ast\alpha}G^{\alpha\beta}\dot\psi_n^\beta +
\dot\psi_n^{\ast\alpha}G^{\alpha\beta}\psi_n^\beta =0 
\;\;. \end{equation}  
For self-adjoint $\hat G$,  defined by Gaussian integers, 
this relation is about real integer quantities.\hfill $\bullet$  

A rearrangement of Eq.\,(\ref{Gconserv}) yields the corresponding 
conserved quantity $q_{\hat G}$ (using matrix notation, as before): 
\begin{equation}\label{qG} 
q_{\hat G}:=\psi_n^*\hat G\psi_{n-1}+\psi_{n-1}^*\hat G\psi_n
=\psi_{n+1}^*\hat G\psi_n+\psi_n^*\hat G\psi_{n+1}   
\;\;, \end{equation}  
{\it i.e.} a real integer-valued two-point correlation function which is invariant 
under a shift $n\rightarrow n+m$, $m\in\mathbf{Z}$. -- In particular, for 
$\hat G:=\hat 1$, the corresponding conservation law amounts to a constraint on 
the state variables: 
\begin{equation}\label{normal}  
q_{\hat 1}=2\mbox{Re}\;\psi_n^*\psi_{n-1}
=2\mbox{Re}\;\psi_{n+1}^*\psi_n=\mbox{const}   
\;\;. \end{equation}  
This can be anticipated to play a similar role for discrete CA as the familiar normalization 
of state vectors in continuum QM.  --   
We may also define the following symmetrized quantity: 
\begin{equation}\label{Q} 
\psi_n^*\hat{\cal Q}\psi_n :=\frac{1}{2}\mbox{Re}\;\psi_n^*(\psi_{n+1}+\psi_{n-1})  
\equiv\frac{1}{2}\mbox{Re}\;\psi_n^{*\alpha}(\psi_{n+1}^\alpha +\psi_{n-1}^\alpha )  
\;\;, \end{equation} 
which, by Eq.\,(\ref{normal}), is conserved as well.   

\subsection{Continuum representation}
We have shown earlier how to construct an one-to-one invertible 
map between the dynamics of discrete Hamiltonian CA and continuum QM in  
presence of a fundamental time scale \cite{PRA2014,EmQM13,Discrete14}.  
Due to this discreteness scale $l$, continuous time wave functions are  
{\it bandlimited}, {\it i.e.}, their Fourier transforms have only finite support 
in frequency space, $\omega\in [-\pi /l,\pi /l]$. 
Hence, we can apply Sampling Theory, in order to reconstruct continuous time signals, 
wave functions $\psi^\alpha (t)$, from their discrete samples, the CA state variables $\psi_n^\alpha$, and {\it vice versa} \cite{Shannon,Jerri,Kempf}. 

Let us represent here the resulting mapping rules obtained 
through the reconstruction formula provided by {\it Shannon's Theorem} \cite{Shannon,Jerri}:    
\begin{eqnarray}\label{psit} 
\psi_n^\alpha &\longmapsto &\;\psi^\alpha (t)
\;\;, \\ [1ex] \label{npm1}
\psi_{n\pm 1}^\alpha &\longmapsto &
\;\exp\big [\mp l\frac{\mbox{d}}{\mbox{d}t}\big ]\psi^\alpha (t)
=\psi^\alpha (t\mp l) 
\;\;, \\ [1ex] \label{samp} 
\psi^\alpha (nl)&\longmapsto &\;\psi_n^\alpha 
\;\;, \end{eqnarray} 
keeping in mind that the continuum wave function is bandlimited. 

These results allow to map CA equations of 
motion, in particular Eqs.\,(\ref{delpsistar})--(\ref{delpsi}) to appropriate 
continuum versions. Corresponding to Eqs.\,(\ref{Gconserv})--(\ref{Q}), there exist 
analogous conservation laws and conserved quantities, which are  
found by applying the mapping rules separately to all wave function factors that appear. 
Thus, for example, we obtain from Eq.\,(\ref{Q}) the conserved quantity: 
\begin{eqnarray}\label{Qcont1}  
\mbox{const}=\psi_n^*\hat{\cal Q}\psi_n&\longmapsto&\; 
\psi^*(t)\hat{\cal Q}\psi (t)
=\mbox{Re}\;\psi^*(t)\cosh\big [l\frac{\mbox{d}}{\mbox{d}t}\big ]\psi (t)
\\ [1ex] \label{Qcont2} 
&\;&=
\psi^{*\alpha}(t)\psi^\alpha (t)
+\frac{l^2}{2}\mbox{Re}\;\psi^{*\alpha}(t)\frac{\mbox{d}^2}{\mbox{d}t^2}\psi^\alpha (t) 
+\mbox{O}(l^4)   
\;\;, \end{eqnarray} 
which shows $l$-dependent corrections to the continuum limit, which here 
amounts to the usual conserved normalization 
$ \psi^{*\alpha}\psi^\alpha =\mbox{const}$\,. 
Similarly, the Schr\"odinger equation with finite-$l$ correction terms are 
obtained \cite{PRA2014}.   

Up to this point, our considerations dealt with individual Hamiltonian CA. 
Based on these results, we will briefly indicate in the following the construction of  
multipartite systems.    

\subsection{From single to multipartite Hamiltonian CA  \cite{Elze16}}  
Can discrete CA combine to form composite multipartite systems? -- 
A positive answer to this question is crucial, if we would like the quantum features of CA that we have found, so far, to also include the full extent of the Superposition Principle. 
It is responsible for interference effects that we can find in single CA. However, a more 
dramatic consequence in QM is the possibility of entanglement  in states or observables.  

When scrutinizing possible constructions of multipartite CA, several requirements appear naturally. -- One may wonder whether not only the {\it linearity} of the evolution law  but 
also the {\it tensor product structure} of composite wave functions finds its analogue here. These are the fundamental ingredients of the usual continuum theory reflected in 
interference and entanglement. Which should be recovered in the continuum limit 
($l\rightarrow 0$) of the CA picture, at least. Furthermore, when the discreteness scale $l$ is finite, the dynamics of composites of CA which do not interact among each other should lead to {\it no spurious correlations} among them. This principle of ``no correlations without interactions'' holds in all of known physics.
 
However, we are facing some serious obstacles which seem to prevent 
satisfying these requirements, when trying to form composites of Hamiltonian CA.  

The want-to-be 
discrete time derivative introduced before, $\dot O_n:=O_{n+1}-O_{n-1}$, for any quantity $O_n$ depending on the clock variable $n$, which appears all over in the 
CA equations of motion and conservation laws, does not obey the   
{\it Leibniz rule}: 
\begin{equation}\label{Leibniz} 
\dot {[A_nB_n]}=
\dot A_n\textstyle{\frac{B_{n+1}+B_{n-1}}{2}}+
\textstyle{\frac{A_{n+1}+A_{n-1}}{2}}\dot B_n 
\neq \dot A_nB_n +A_n\dot B_n 
\;\;. \end{equation} 
Similar observations can be expected for other definitions one might come up with. 
Ignoring this for a moment,  
consider a trial multi-CA equation of motion analogous to the single-CA 
Eq.\,(\ref{delpsistar}): 
\begin{equation}\label{PSIeq} 
\dot\Psi_n=\frac{1}{i}\hat H_0\Psi_n 
\;\;, \end{equation} 
where $\hat H_0$  may describe a block-diagonal Hamiltonian in the absence of 
interactions among the CA. Then, through Eq.\,(\ref{Leibniz}),  the 
required {\it factorization} 
of Eq.\,(\ref{PSIeq}) is hindered on the left-hand side, since unphysical correlations  
will be produced among the components of a factorized wave function, such as   
\begin{equation}\label{PSI} 
\Psi_n^{\alpha\beta\gamma\cdots}=
\psi_n^\alpha\phi_n^\beta\kappa_n^\gamma\cdots 
\;\;, \end{equation}  
or for superpositions of such factorized terms. 

For a bipartite system, for example, we obtain immediately:  
$\dot \Psi_n^{\alpha\beta}=
\dot\psi_n^\alpha (\phi_{n+1}^\beta+\phi_{n-1}^\beta )/2+ 
\psi\leftrightarrow\phi 
\;\neq\;\dot\psi_n^\alpha\phi_n^\beta +\psi_n^\alpha\dot\phi_n^\beta$. -- 
Furthermore, applying the mapping rules of Section\,2.3, before taking the 
limit $l\rightarrow 0$, we find that the bilinear terms here do not 
converge to the appropriate QM expression.  
Which should be   
$\partial_t(\psi^\alpha\phi^\beta )=(\partial_t\psi^\alpha )\phi^\beta +\psi^\alpha \partial_t\phi^\beta$, in order to allow     
the decoupling of two subsystems that do not interact.  

This latter problem is a general one of nonlinear terms in the equations 
of motion of discrete CA, 
which we discussed before \cite{Discrete14}:  {\it The linear map provided by 
Shannon's Theorem does not commute with the multiplication implied by the 
nonlinearities.}  This follows from the 
explicit reconstruction formula (or any variant thereof that is linear) \cite{PRA2014,Shannon,Jerri}.  
Also on the right-hand side of Eq.\,(\ref{PSIeq}) we find this obstruction, 
when trying to map the equation with a factorized wave function to its continuous 
time description.     

\subsubsection{The many-time formulation} 
It can be observed that 
the  difficulties we pointed out arise from the implicit assumption that the components of a multipartite CA are {\it synchronized} to the extent that they share a common clock 
variable $n$. In Ref.\,\cite{Elze16}, we have considered a radical way out of the impasse 
encountered 
by resorting to a {\it many-time} formalism. This means giving up synchronization 
among parts of the composite CA by introducing a set of clock variables, 
$\{ n(1),\;\dots ,\;n(m)\}$, one for each one out of $m$ components.  

It may be surprising to encounter this in the present nonrelativistic context, since the 
many-time formalism has been introduced by Dirac, Tomonaga, and Schwinger in 
their respective formulations of relativistically covariant many-particle 
QM or quantum field theory, where a global synchronization cannot be 
maintained \cite{Dirac,Tomonaga,Schwinger}.   
  
We replace here the single-CA action of Eq.\,(\ref{action}) by 
an integer-valued multipartite-CA action: 
\begin{equation}\label{maction} 
{\cal S}[\Psi ,\Psi^*] :=\Psi^*\big (
\sum_{k=1}^m\hat{\cal S}_{(k)}\;+\;\hat{\cal I}\big )\Psi
\;\;, \end{equation} 
with $\Psi :=\Psi^{\alpha_1\dots\alpha_m}_{n_1\dots n_m}$ and, correspondingly,  $\Psi^*$ as independent {\it Gaussian integer} variables; 
the self-adjoint operator $\hat{\cal I}$ incorporates interactions between different CA; 
whereas
$\hat{\cal S}_{(k)}$ is as introduced in Eq.\,(\ref{action}), with the subscript 
$_{(k)}$ indicating that it acts {\it exclusively} on the pair of indices pertaining to 
the $k$-th single-CA subsystem:   
\begin{equation}\label{Sopk} 
\Psi^*\hat{\cal S}_{(k)}\Psi:=
\sum_{\{ n_k\} }\big[ (\mbox{Im}\;
\Psi_{\dots n_{k}\dots}^{*\dots\alpha_{k}\dots}
\;\dot\Psi_{\dots n_k\dots}^{\dots\alpha_k\dots}
\;+\;\Psi_{\dots n_{k}\dots}^{*\dots\alpha_{k},\dots}\;
H_{(k)}^{\alpha_k\beta_k}\Psi_{\dots n_k\dots}^{\dots\beta_k\dots}
\big ]
\;\;, \end{equation} 
with summation over {\it all} clock variables (summation over 
twice appearing Greek 
indices remains understood); the $\dot{\phantom .}$-operation, however, acts only with 
respect to the explicitly indicated $n_k$, $\dot f(n_k):=f(n_k+1)-f(n_k-1)$, while the 
single-CA Hamiltonian, $\hat H_{(k)}$, requires a matrix multiplication, as before.  

Obviously, we can apply the {\it CA Action Principle} also to the present situation with 
the generalized action of Eq.\,(\ref{maction}). This results in the following discrete equations of motion: 
\begin{equation}\label{mEoM} 
\sum_{k=1}^m\dot\Psi_{\dots n_k\dots}^{\dots\alpha_k\dots}
\;=\;\frac{1}{i}\big (\sum_{k=1}^mH_{(k)}^{\alpha_k\beta_k}
\Psi_{\dots n_k\dots}^{\dots\beta_k\dots} 
\;+\;{\cal I}^{\dots\alpha_k\dots\;\beta_1\dots\beta_m}
\Psi_{\dots n_k\dots}^{\beta_1\dots\beta_m} 
\big )
\;\;, \end{equation}  
together with the adjoint equations; here the interaction $\hat{\cal I}$, 
like $\hat H_{(k)}$, is assumed to be independent of the clock variables and    
the $\dot{\phantom .}$-operation acts only with respect to 
$n_k$ in the $k$-th term on the left-hand side.   

We have verified in Ref.\,\cite{Elze16} that the many-time formulation avoids  
the problems of a single-time multi-CA equation, such as Eq.\,(\ref{PSIeq}), which we 
discussed. 

In particular, in the absence of interactions with each other,   
between CA subsystems, $\hat{\cal I}\equiv 0$, {\it no unphysical correlations} are introduced among independent CA subsystems. 

Furthermore,    
{\it continuous multi-time equations} corresponding to Eqs.\,(\ref{mEoM}) are obtained 
by applying the mapping rules given in Section\,2.3. to the discrete equations.   
We find no problem of incompatibility between multiplication according to nonlinear 
terms {\it vs.} linear mapping  according to 
{\it Shannon's Theorem}, since a separate mapping is applied for each one of the 
clock variables. This effectively replaces $n_k\rightarrow t_k,\;k=1,\dots,m$, 
where $t_k$ is a continuous real time variable.  In this way, a {\it modified 
multi-time Schr\"odinger equation} is obtained:  
\begin{equation}\label{mSchroed} 
\sum_{k=1}^m\sinh\big [l\frac{\mbox{d}}{\mbox{d}t_k}\big ] \Psi_{\dots t_k\dots}^{\dots\alpha_k\dots}
\;=\;\frac{1}{i}\big (\sum_{k=1}^mH_{(k)}^{\alpha_k\beta_k}
\Psi_{\dots t_k\dots}^{\dots\beta_k\dots} 
\;+\;{\cal I}^{\dots\alpha_k\dots\;\beta_1\dots\beta_m}
\Psi_{\dots t_k\dots}^{\beta_1\dots\beta_m} 
\big )
\;\;, \end{equation}  
where an overall factor of two from the left-hand side has been absorbed into the matrices  on the right. By construction, here $\Psi$ is bandlimited  
with respect to each variable $t_k$. 

Performing the continuum limit, $l\rightarrow 0$, we arrive at the  
multi-time Schr\"odinger equation (one power of $l^{-1}$ providing the 
physical dimension of $\hat H_{(k)}$ and $\hat {\cal I}$) considered by Dirac and 
Tomonaga \cite{Dirac,Tomonaga}. However, when $l$ is fixed and finite, modifications 
in the form of powers of $l\mbox{d}/\mbox{d}t_k$ arise on its left-hand side, 
similarly as in single CA case before.     

In the present nonrelativistic context, it may be appropriate to identify  
$t_k\equiv t,\;k=1,...,m$, in which case the operator on the left-hand side of 
Eq.\,(\ref{mSchroed}), for $l\rightarrow 0$, can be simply replaced by 
$\mbox{d}/\mbox{d}t$. This results in the usual (single-time) 
{\it many-body Schr\"odinger equation}.  
  
Finally, the study of the {\it conservation laws} of the multipartite CA equations of motion 
can be performed along the lines of Section\,2.2. and analogous results have been 
obtained \cite{Elze16}. 

\subsubsection{Remarks on the Superposition Principle in composite CA}   
The equivalent discrete or continuous many-time equations (\ref{mEoM}) and 
(\ref{mSchroed}) are both linear in the CA wave function $\Psi$. Therefore, superpositions 
of solutions of these equations also present solutions and the 
{\it Superposition Principle} does indeed hold for multipartite Hamiltonian 
CA.    

As in the case of single CA, this entails the fact that these discrete systems 
 -- with all variables, parameters, {\it etc.} presented by Gaussian integers -- can 
produce {\it interference} effects as in quantum mechanics. Even more interesting, their 
composites can also show {\it entanglement}, which is deemed an essential feature of QM.    
This follows from the form of the equations of motion, which allow for superpositions 
of factorized states. 

For example, in the bipartite case 
($k=1,2$), assuming that the individual CA are characterized by two degrees of freedom 
($\alpha_k=0,1$), a time dependent analogue of the totally antisymmetric   
{\it Bell state} is given by: 
\begin{equation}\label{Bell} 
\Psi\propto\psi_{n_1}^{\alpha_1=0}\psi_{n_2}^{\alpha_2=1}
-\psi_{n_1}^{\alpha_1=1}\psi_{n_2}^{\alpha_2=0} 
\;\;, \end{equation} 
which may be a solution of appropriate discrete equations of motion. 
  
We conclude with a word of warning concerning the freely used expressions here that we borrowed from QM, such as ``wave functions'' and ``states'', in particular. They usually  
invoke the notion of vectors in a {\it Hilbert space}, which turns into a complex 
projective space upon normalization of the vectors. However, 
as has become obvious in Section\,2.2., 
see Eqs.\,(\ref{normal})--(\ref{Q}), and which can be seen similarly in the multipartite 
CA case \cite{Elze16}, as 
long as the CA are truly discrete ($l\neq 0$), the normalization (squared) of vectors is not among the conserved quantities, hence not applicable, but is replaced by a conserved 
two-time correlation function instead. 

Furthermore, contrary to first impression, perhaps, 
the envisaged space of states is {\it not} a 
Hilbert space, since it fails in two respects: the vector-space and completeness 
properties are missing.  --  First of all, 
the relevant {\it Gaussian integers} are not 
complete. Hence the completeness property of the space of states is lacking, which is 
built with these integers as underlying scalars. Secondly, the integers {\it only} 
featuring in all aspects of the CA do not form a 
field but a commutative ring (for the multiplication of vectors by such scalars 
there is no multiplicative inverse, such as exists, {\it e.g.}, for rational,  
real, or complex numbers). Therefore, we do not have a vector space over a field, as 
usual in QM. It has to be replaced  by a more general structure. This is known as a 
module over a  ring, presently a module over the commutative ring of 
Gaussian integers. This allows the construction of a linear space endowed with an 
integer-valued scalar product, {\it i.e.} a {\it unitary space}. Taking its incompleteness 
into account, then, the space of states in the presented CA theory can be classified as a 
{\it pre-Hilbert module over the commutative ring of Gaussian integers}. 

In any case, we conclude that superpositions of states, interference effects, and entanglement, as in QM, all can be found already on the ``primitive'' level 
of the presently considered natural Hamiltonian CA, discrete single or multipartite systems which are characterized by (complex) integer-valued variables and couplings.    

\section{Conclusion} 
This presents a brief review of earlier work which has demonstrated surprising quantum features arising in integer-valued, hence ``natural'', {\it Hamiltonian cellular 
automata} \cite{PRA2014,EmQM13,Discrete14,Elze16,WignerSymp13}.  

The study of this 
particular class of CA is motivated by 't\,Hooft's {\it Cellular Automaton 
Interpretation of QM} \cite{tHooft2014} and various recent attempts to construct models 
which may eventually lead to demonstrating that the essential features of QM can 
all be understood to emerge from pre-quantum deterministic dynamics.    

The single CA we have considered allow practically for the first time to 
reconstruct quantum mechanical models  with nontrivial Hamiltonians in terms of 
such systems with a finite discreteness scale. -- 
Furthermore,  we have recently extended 
this study by describing {\it multipartite systems}, analogous to many-body QM. 
Not only is this useful for the construction of more complex models {\it per se} 
(especially with 
a richer structure of energy spectra), but it is also necessary, in order to  
extend the {\it Superposition Principle} of QM to a description at the CA level. 
We find that it can be introduced already there to the fullest extent, 
compatible with a tensor product structure of multipartite states, 
which entails the possibilities of their {\it interference} and {\it entanglement}. 

Surprisingly, we have been forced -- in our approach employing Sampling Theory for the 
map between CA and an equivalent continuum picture -- to introduce a 
many-time formulation, which only appeared in relativistic quantum mechanics before,  
as introduced by Dirac, Tomonaga, and Schwinger \cite{Dirac,Tomonaga,Schwinger}.  
This points towards a crucial further step in these developments, which is still missing, 
namely a CA model of {\it interacting quantum fields}. Without the possibility of  
interacting multipartite CA with quantumlike features, as described here, it is hard to 
envisage a CA picture of dynamical fields spread out in spacetime.   
Last not least, a detailed model is still lacking in which the Born rule can be 
related quantitatively to the ``ordinary'' statistical mechanics of 
deterministic dynamical degrees of freedom.   

\ack
It is a pleasure to thank G.\,Gr\"ossing and J.\,Walleczek for the kind invitation to the conference ``Emergent Quantum Mechanics'' (Vienna, October 2015) and H.\,Batelaan, 
A.\,Caticha and G.\,'t~Hooft for discussions.  Support by the Fetzer-Franklin Fund is gratefully acknowledged.  


\end{document}